# Observing periodic gap variations in cuprates

Riju Banerjee[1,3*†], Emily L. Wang[2,3*], Eric W. Hudson[3]

1. National Physical Laboratory, Teddington, TW11 0LW, United Kingdom

2. Department of Computer Science, Northwestern University, Evanston, IL 60201, USA

3. Department of Physics, The Pennsylvania State University, University Park, Pennsylvania 16802, USA

* equal contributions, †email: riju.banerjee@npl.co.uk

**Abstract**

Central to the enigma of the cuprates is ubiquitous electronic inhomogeneity arising from a variety of electronic orders that coexist with superconductivity, the individual signatures of which have been impossible to disentangle despite four decades of intense research. This strong nanoscale inhomogeneity complicates interpretation of measurements both by probes which average over this inhomogeneity and those, like scanning tunneling microscopy (STM), which should be able to spatially resolve variations driven by both order and inhomogeneity. Here, we develop a novel technique that directly acknowledges this electronic inhomogeneity and extracts statistically significant features from scanning tunneling spectroscopic data. Applying our novel technique to single and bilayer Bi-based cuprates spanning a large doping range, we peer through local inhomogeneities and find that the gap breaks translational and rotational symmetries and varies periodically in a four-fold pattern. Our direct observation of a symmetry breaking gap in the single particle tunneling spectra adds strong credence to the pair density wave hypotheses supposed to exist in these materials. We also discuss various implications of our observations and, in particular, how they can explain the origin of the low energy checkerboard pattern.

**Introduction:**

Materials with strong electronic correlations often feature a plethora of electronic orders that coexist or compete with each other. Strong electron-electron interactions, acting through spin, charge, orbital or lattice degrees of freedom, result in a highly inhomogeneous electronic structure experimentally found to be ubiquitous in these materials [1]. For example, one of the most heavily studied strongly correlated system, the high temperature superconductor cuprates, have been shown to exhibit signatures of electronic phase separation [2], nematicity [3,4], checkerboard patterns [5,6] and other local order [7],while also exhibiting local inhomogeneities, perhaps seeded by disorder [8], in transition temperature $T_c$ [9], particle-hole asymmetry [10] and Fermi surface [11]. Unfortunately, to date, it has been impossible to find a consistent theoretical description encompassing all the intertwined orders in these materials. Indeed, theoretically modelling this omnipresent inhomogeneous electronic order is so challenging that it is usually simply ignored.

Some of the most significant evidence of inhomogeneity in cuprates has been gathered through scanning tunneling microscopy (STM). The single particle tunneling spectrum (Fig. 1a) is often reminiscent of what would be expected for a d-wave superconductor, with a v-shaped gap, symmetric about the Fermi-energy, and peaked at the maximum gap energy $\Delta$. Extracting this peak energy, either algorithmically [2] or via some fitting technique [12,13] is a common first analysis step, and a spatial map of the energy of this feature (a "gap-map", as shown in Fig. 1b) further highlights the inhomogeneity in these materials. However, other spectral features besides this peak also exist, including higher energy features, which are usually extremely challenging to measure, but were found to be tied closely to non-stoichiometric Oxygen atom dopants [8], and a lower energy "kink," (Fig. 1a) often hidden by the more dominant peak feature. More than a decade ago, one of us (EWH) argued [14] the peak energy to be associated with the pseudogap (a non-superconducting order in the material) and the kink with superconductivity, and these two spectroscopic features are often labeled accordingly as $\Delta_{PG}$ and $\Delta_{SC}$, even though such an interpretation is still somewhat debated [15,16].

Interestingly, within this strongly spatially inhomogeneous single-particle spectrum, both uniformity and order have also been observed. The same spectral surveys (measurements of the energy, $E$, and position, $\vec{r}$, dependent differential tunneling conductance $g(E,\vec{r})$) which yield highly inhomogeneous gap maps also reveal, through Quasiparticle interference (QPI) studies [17], that the Bogoliubov quasiparticles responsible for superconductivity are well defined and coherent throughout the sample [18]. In contrast to the apparently global orders revealed by QPI, at the level of individual spectra, this coherence of the Bogoliubov quasiparticles (and hence the superconducting order) is also suggested by the uniformity of the kink energy [14], which remains constant defying all underlying electronic inhomogeneity.

These hints of homogeneity are rare, but they confirm the existence of persistent patterns in the cuprate electronic structure, obscured behind the façade of inhomogeneity. Given that, it is perhaps surprising that after several decades of study and multiple theoretical predictions, cuprate gap maps have not revealed any order themselves. For example, pair density wave (PDW) models, built on a decades old theoretical paradigm that has recently gathered considerable experimental support, specifically predict spatial, periodic gap variation [19–24]. Here, we introduce a novel technique to disentangle the gap (spectral peak) energy from underlying inhomogeneities, and find that the spectral peak does indeed vary in a four-fold symmetric pattern and that its interplay with the kink is responsible for the appearance of the checkerboard pattern [25,26] at low energies.

**Main body:**

**The Statistically Unique Feature Finder**

Conventional gap-finding algorithms analyze each spectrum individually, either algorithmically searching for sudden changes in density of states (peaks) or slopes to identify the gap edge, or by using a curve fitting technique [2,12,13]. However, such a strategy can be negatively impacted by inhomogeneity, particularly if local dopants or defects and various orders contribute to the electronic structure. To get around this issue, we have developed the STatistically Unique Feature Finder (STUFF), which compares each spectrum to others nearby (within a few nanometers), in order to find spectral features as they appear given the local background.

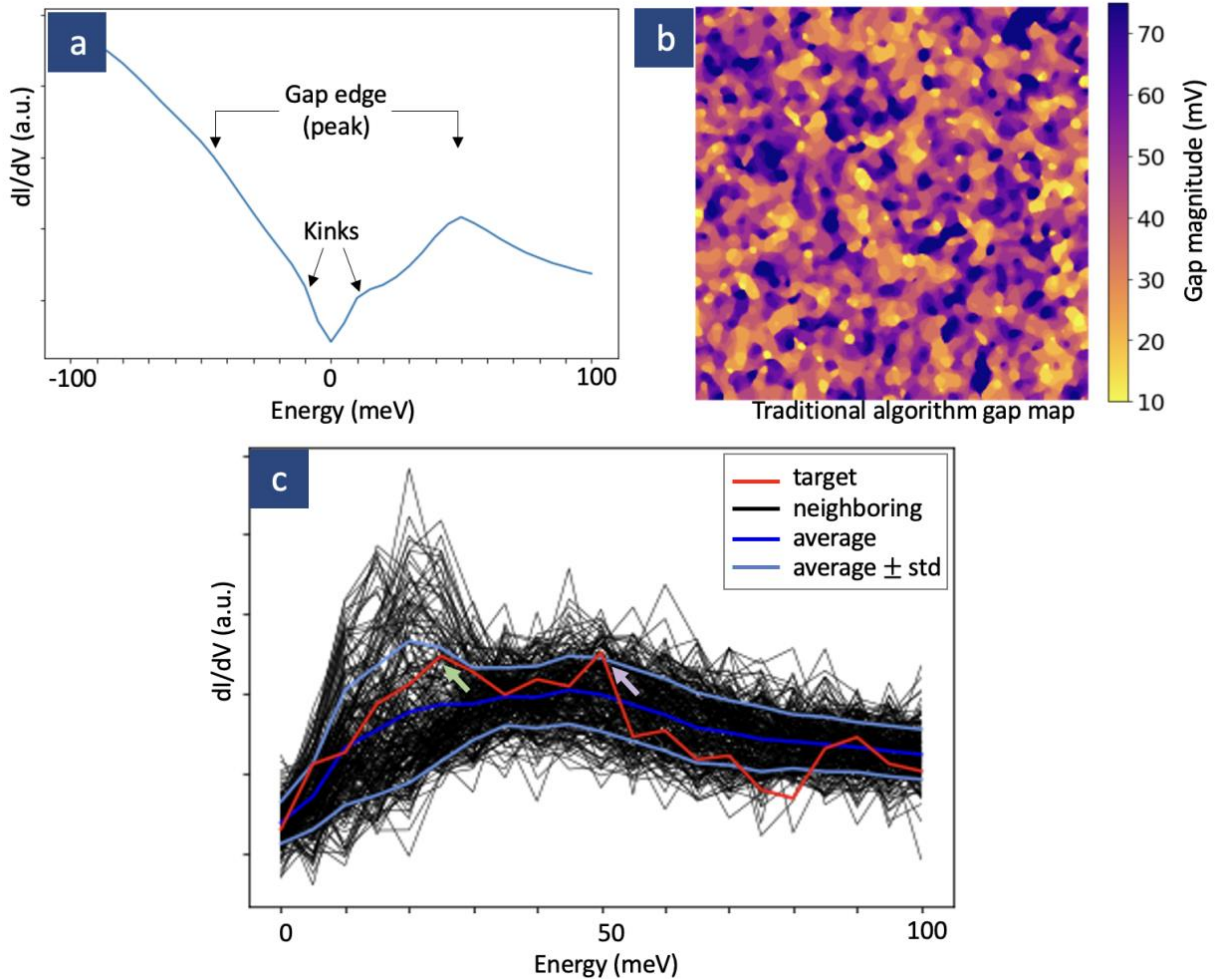

Figure 1: (a) Spatially averaged single particle tunneling spectrum obtained from the dI/dV conductance maps over a 68.5 X 68.5 nm$^2$ area broken into 400 X 400 pixels on a region of underdoped Bi-2201 ($T_c$=32 K). Typical spectral features like the gap, kinks and the gap edge (peaks) are marked. (b) For the same dataset from which (a) was obtained, the traditional "gap" (peak) map algorithm considers each spectrum individually and shows strong local electronic variation without any clear order. (c) The idea behind Statistically Unique Feature Finder (STUFF) algorithm. Each target spectrum (red curve) is compared to approximately two hundred spectra around it (black curves) to identify a z-score (Equation 1). For the target spectrum in red, the peak at 25 meV (marked by green arrow) can naively be identified as the gap edge. However, that feature is common to many other neighborhood spectra, suggesting that it might be arising from a local (unidentified) electronic order. Instead, STUFF identifies the peak at 50 meV (marked by violet arrow) as the gap edge as that is a peak feature statistically most unique to the target spectrum. When the algorithm is executed for each spectrum in the field of view, the new gap map obtained from STUFF is shown in Fig. 2b.

The idea behind STUFF is demonstrated in Fig. 1c, where it is applied on an underdoped Bi-2201 ($T_c$=32K) sample. The sample also has Pb dopants to reduce structural supermodulations. Considering the (arbitrarily chosen) target spectrum shown in red, we see that there is a peak near 25 meV (only positive bias half is shown), which is identified by the traditional algorithm as the gap edge. However, considering more than two hundred spectra obtained in a 68.5 X 68.5 nm$^2$ neighborhood (shown in black), we find they all have a comparable peak-like feature around the same energy. This suggests that this particular feature might be arising from an unidentified background order. Examining the target spectrum, we see that there is another distinct peak-like feature around 50 meV, which is absent in other spectra taken from the same region. STUFF thus identifies this peak energy, as it is a statistically unique peak feature specific to the target spectrum.

In practice, STUFF is implemented by calculating a z-score for each spectrum $g(E)$ as

$$z(E) = \frac{g(E) - \bar{g}(E)}{\sigma(E)} \quad (1)$$

where $\bar{g}(E)$ is the average spectrum of the neighborhood (shown in dark blue in Fig. 1c) and $\sigma(E) = \sqrt{\frac{1}{N}\sum_{i=1}^{N}[g_i(E) - \bar{g}(E)]^2}$ is the standard deviation of all *N* spectra in its neighborhood, describing the local variation at each energy. We note that for the target spectrum (red in Figure 1c), the z-score will be highest for the feature at 50 meV (marked by violet arrow), consistent with the observation above. Thus, the STUFF algorithm selects out peak-like features unique to the target spectrum by comparing it to its neighboring spectra. We run the STUFF algorithm for every spectrum in the field of view (FOV) and determine the energy for which the z-score is the highest for each spectrum. To preserve consistency with previous work, we will refer to this identified peak energy (for each spectrum, the energy for which the z-score is highest) as "the gap" $\Delta_{ST}$, and note that it is very similar to "the pseudogap energy" $\Delta_{PG}$ determined by conventional techniques. Finally, we use this identified peak energy for each spectrum to redraw the gap map.

**Results from STUFF**

We apply this technique to two different cuprates, underdoped Bi-2201 (T$_c$ = 32 K) and overdoped Bi-2212 (T$_c$ = 75 K). Both samples are doped with Pb to reduce the structural supermodulation that exists in the undoped materials and can also result in periodic gap variations [27,28] (Fig. 2a). The samples were studied at 4K using a custom-built variable-temperature scanning tunneling microscope. The spectral surveys were corrected for thermal drift [29]. Our analysis reveals that for both samples the gap varies spatially in a four-fold symmetric pattern (Fig. 2b), which was not evident using the conventional gap-finding algorithm (Fig. 1b). Average spectra for each gap value are shown in Fig. 2c using the same color indexing scheme used in Fig. 2b. The fanning out of these spectra, with gaps from 10 meV to 65 meV, is essentially indistinguishable from similar analysis using the conventional algorithm [5,14], making the dramatic difference in the spatial gap map even more remarkable.

From Eq. 1, it follows that the only free parameter in our technique is the size of the neighborhood we consider in calculating the z-score (varying the number of neighborhood spectra, *N*). We next check that our observation of periodic gap variation is not an artifact of partitioning the spectral map into these smaller regions. The effect of changing the neighborhood size is demonstrated in Fig. 2d-f. Considering larger and larger neighborhoods, we see that the length scale of gap variation is insensitive to the size of the neighborhood. Remarkably however, as the neighborhood size is increased, and each spectrum is compared to a larger number of spectra around it, identifying the gap like feature becomes harder and harder, causing the four-fold symmetric pattern to gradually become weaker. When the neighborhood is increased to the entire field of view, the gap map from STUFF approaches the traditional gap map algorithm shown in Fig. 1b. In supplementary section 1, we present a similar analysis on another sample, Bi-2212 ($T_c$=75K) and make similar observations. We also note that the four-fold symmetric pattern looks similar to the well-known checkerboard pattern (Fig. 2g), hinting at a possible relation between the two, which we address later in the paper.

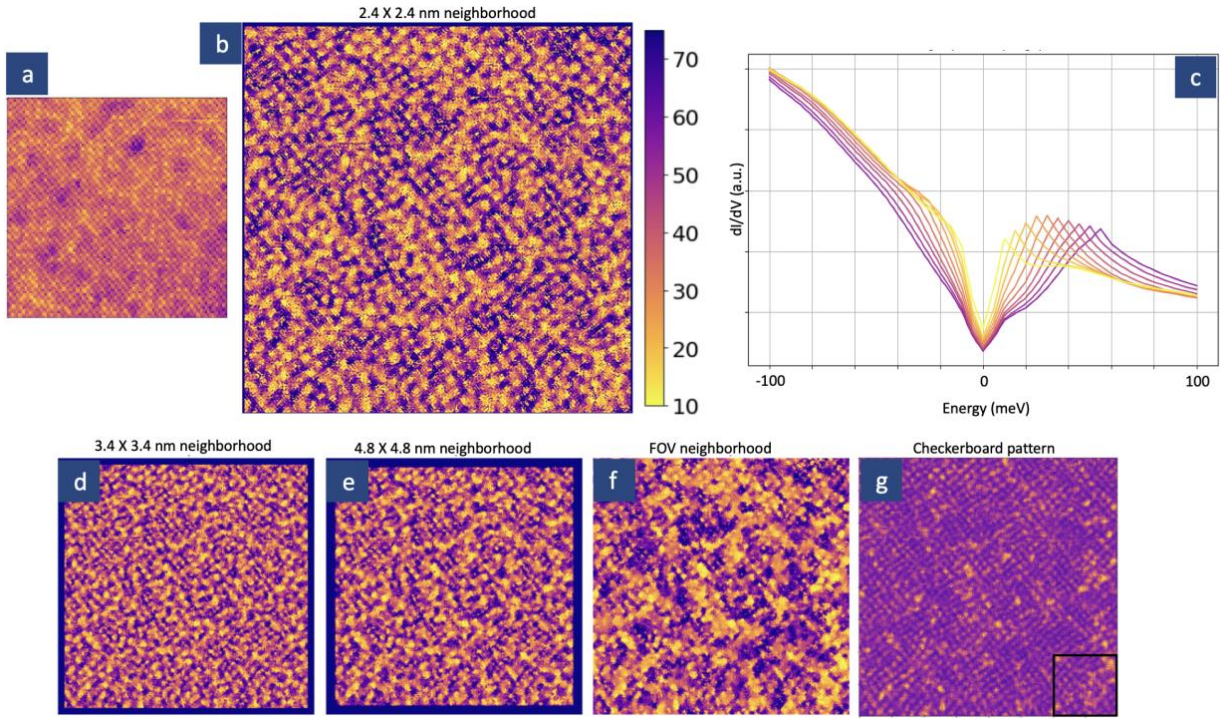

Figure 2: (a) Atomic resolution topography shows the lack of any supermodulation in the Pb doped Bi-2201 ($T_c$=32K) sample. (b) Redrawn gap map obtained from our STUFF algorithm using a 2.4 X 2.4 nm (14 pixels square) neighborhood. Note the emergence of a clear four-fold symmetric pattern, in contrast to the lack of any order detected by the traditional algorithm (Fig. 1b), with the spatially average spectra corresponding to the colors plotted in (c). The color bar is shared across images (b-g) and is the same as in Fig. 1b. (d-f) Increasing the neighborhood size considered for calculating the z-score does not change the change the periodicity of the four-fold symmetric pattern, but the pattern gets blurred and gradually morphs towards the traditional gap map (Fig. 1b). (g) 10 meV layer of the spectral map, demonstrating four-fold symmetry of the checkerboard pattern. The black box marks the region of the atomic resolution topography in (a).

The fact that the length scale of gap variation is not an artifact of our analysis is further demonstrated in Fig. 3a-d where we show FFTs of the gap maps in Figure 2d-g. Line cuts taken through the symmetric axis (marked by the black line in Fig. 3a-d) demonstrate that increasing the neighborhood only reduces the contrast of the four-fold symmetric pattern, without changing its periodicity. We also note the progressive increase in weight in in the center of the FFT pattern in Fig. 3a-d. The center of the FFT corresponds to a-periodic features (noise), and that increases as STUFF's efficiency is reduced.

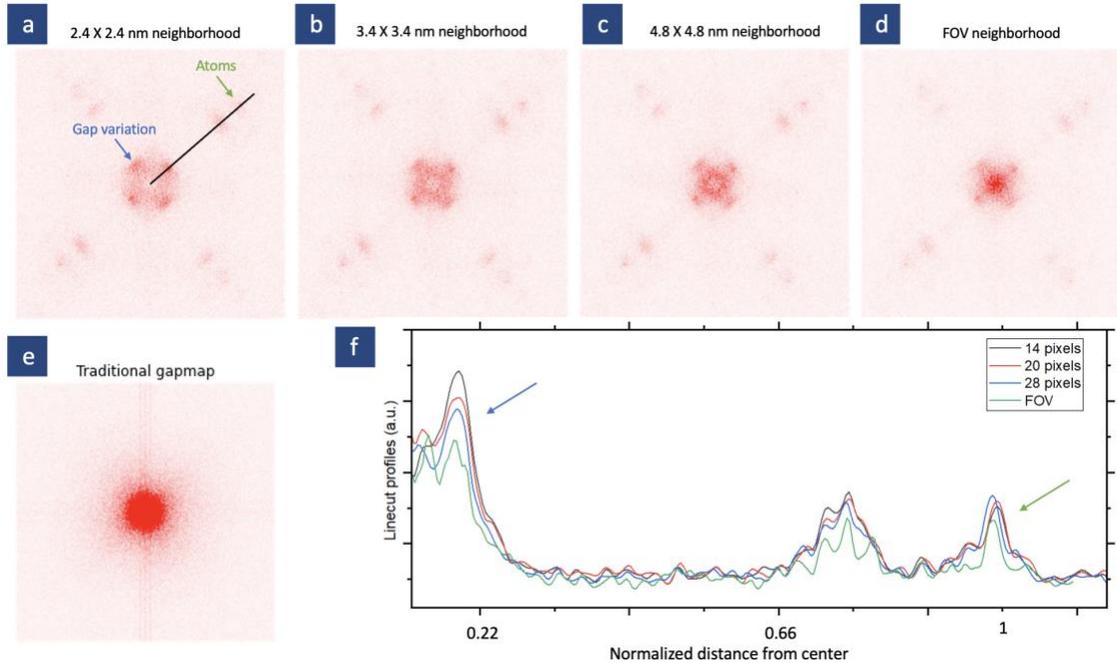

Figure 3: FFTs of the gap maps obtained from STUFF and shown in Fig. 2. (a-d) A clear four-fold symmetric gap variation pattern emerges using our STUFF algorithm, whose periodicity is marked by blue arrow. In contrast, the FFT of the traditional gap map, shown in (e), has no periodicity. STUFF is also able to pick out the atomic periodicity (green arrow). Note that observed periodicity of the four-fold symmetric gap variation pattern does not depend on the size of the neighborhood considered for running the STUFF algorithm, implying that the periodicity is inherent to the sample, and not an artifact of the algorithm. However, increasing the neighborhood size does reduce STUFF's efficacy, as seen in the progressive transfer of weight from the peaks to the center, leading to a reduction in peak height shown in (f).

We next note that even though STUFF reveals a clear four-fold symmetric gap variation, the size of the gap still varies considerably throughout the sample, similar to the traditional gap map algorithm. This is manifested as local variations in contrast in Fig. 2b. We now turn to understanding how the gap changes as a function of the local strength of the four-fold symmetric wave. Identifying the peaks in the four-fold symmetric pattern (Fig. 2b), we break the field of view into Voronoi cells and color (bin) them in five different groups depending on the gap size near the center of each cell (Fig. 4a). The groups are colored blue, orange, green, red and violet respectively in order of increasing gap size at the cell center. Further subdividing each Voronoi cell into six concentric spatial regions, we show intra cell gap variations in Fig. 4b. Cells with smaller gaps near the center (in blue) show greater intra cell gap variation than cells with larger gaps in their center (violet). In Fig. 4c and d, we plot the average spectra for the different distance groups showing this intra cell gap variation. Plots with light to dark show the variation from the cell center to its boundary. Note that the spectra farther away from the cell centers (dark blue and dark purple in Fig. 4c and d respectively) are similar. This implies that the electronic distribution in the sample has a large gap 'sea' which is broken up into 'island' regions with smaller gaps arranged in a fourfold symmetric pattern. This picture is similar to an electronic phase separation idea hypothesized decades ago [2]. Further research would be required to understand what creates the small gap 'islands' and how they remain segregated from the large gap 'sea' regions.

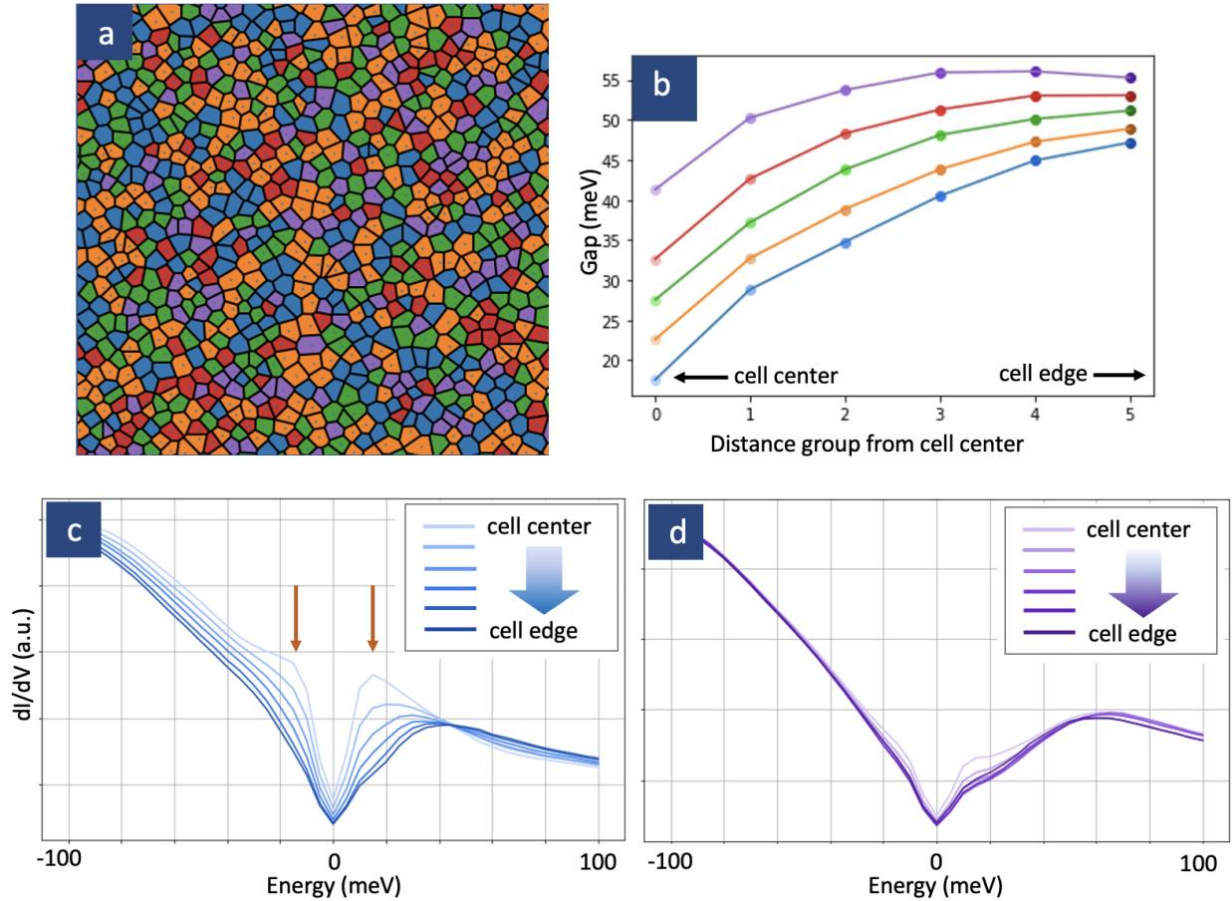

Figure 4: (a) To study how the wave affects the gaps locally, we divide the field of view into Voronoi cells, centered on peaks associated with the fourfold symmetric pattern, and bin them into five different groups depending on the gap magnitude at the cell center. As shown in Fig. 2, the cells have smaller gaps near the center and larger gaps towards their edges. The colors blue, orange, green, red and violet indicate the gap size at the cell center in increasing order. (b) More gap variation is seen in cells with smaller gaps near the center. This implies a 'sea' of large gaps being broken into four-fold symmetric 'islands' of smaller gapped regions. (c) and (d) show the average spectra of the Voronoi cells with smallest (colored blue) and largest (colored violet) gaps near the centers. Note that in (c), for 'islands' with small gaps, significantly prominent peaks are seen (marked by brown arrows) compared to 'islands' with large gaps in (d).

**Discussions and Conclusions:**

**Interpretations and relation to PDW and pseudogap:**

Amidst all the caveats concerning various aspects of cuprates, it is generally agreed that there are two electronic phases that occur simultaneously, namely superconductivity and pseudogap. Understanding the origin of these two phases and how they compete, cooperate, or merely coexist with each other has been the subject of intense debate over the past several decades. Recent experimental evidence [19–24] has suggested that the physics of cuprates and other high temperature superconductors can be explained by pair density waves (PDW). In fact, several theoretical studies have also hypothesized that PDW is the 'parent order' in these materials, from which, other orders like pseudogap and the checkerboard order naturally emerge [30–32]. As our observation of a periodic gap variation aligns well with the predictions of bidirectional PDW, we next turn to examining our observations in light of different proposed pseudogap and PDW theories.

We note that our observation of a four-fold symmetric gap map looks very similar to the checkerboard pattern, which arises from the periodic variation of the spectral weight (density of states) at low energies. One of the earliest attempts to explain the checkerboard pattern was using a spatially varying gap arising from PDW [19,21,33–35]. We next argue that the periodic gap variation obtained from STUFF results in a corresponding periodic variation of the low energy spectral weight and can thus explain the origin of the checkerboard pattern. In plotting the group average spectra obtained from our STUFF algorithm in Fig. 2c, we see that all of them have a distinct kink at 10 meV. This observation suggests that there are in fact two distinct gaps, originating from the superconducting gap ($\Delta_{SC}$, visible as a kink and which remains constant throughout) and the pseudogap ($\Delta_{PG}$, which varies periodically) phases. As the STUFF algorithm selects out only the locally varying parameter, ignoring local uniformities, we identify the STUFF gap $\Delta_{ST} = \Delta_{PG}$. Fig. 4 demonstrates how close these two gaps are in energy and how they interact with each other. In regions where $\Delta_{PG} \gg \Delta_{SC}$, the two gaps are clearly distinguishable in the average spectra (Fig. 4d). In contrast, in regions where $\Delta_{PG} \approx \Delta_{SC}$, the interaction between the two energy scales results in high coherence peaks (Fig. 4c) and increased spectral weight in the gap in both Bi-2201 and Bi-2212 samples (see supplementary section 1 for Bi-2212 data). It is then straightforward to realize that the periodic gap variation can result in periodic low-bias conductance variation, resulting in a checkerboard pattern. Thus, the checkerboard pattern is but a low energy manifestation of a spatially varying pseudogap, which varies over a much wider energy range. Our observations likely also explain why the checkerboard pattern is much easier to observe directly in STS than the gap variation itself. This is likely because all the variation of $\Delta_{PG}$ and inhomogeneities must necessarily stay sufficiently outside of the superconducting gap (leaving it uniform) to have a well-defined gap everywhere and hence a superconductor.

In the above discussion, we argued that the spatially varying gap observed from STUFF is related to the pseudogap $\Delta_{PG}$ and showed how it interacts with the spatially uniform superconducting gap, $\Delta_{SC}$, to create the checkerboard pattern. Quantum oscillations (like the checkerboard) in the pseudogap phase is an integral part of cuprate physics [36] and many studies have tried to relate them to PDW [21,30,31,35,37,37–43]. Our observation of a spatially varying pseudogap adds further support to such hypotheses. Another interesting aspect of the pseudogap phase that is generally agreed upon is that the pseudogap strength goes down with increasing doping level, even though the exact pseudogap termination point remains debatable. In supplementary section 2, we present the result of STUFF algorithm on two more Bi-2201 samples with higher hole doping. We show that compared to the underdoped ($T_c$=32 K) sample, the gap variation pattern is significantly weakened for the optimally doped ($T_c$=35K, Fig. S2) sample and almost imperceptible for the overdoped ($T_c$=15K, Fig. S3) sample. This weakening of the pattern suggests that the spectral variation observed is related to the pseudogap phase, and is consistent with theoretical proposals arguing the existence of a quantum critical point separating an ordered pseudogap and a disordered phase [44–46].

Although a PDW for cuprates was originally hypothesized decades ago [21], it has gathered considerable experimental support in recent years [30,35,39–41,47]. To date, several experiments like the onset of c-axis superconductivity in LBCO at temperatures far below Tc [48]; spatially periodic variation of Cooper pair density observed by scanning Josephson tunneling microscopy [49]; and halving of the charge density wave vector inside vortex halos under magnetic fields [50], have provided considerable, albeit circumstantial evidence for PDW in cuprates and unfortunately only in systems at a single doping level. Due to their strong spatial inhomogeneity, direct observation of gap variation using single particle tunneling in strongly correlated materials is rare [51], and has only been possible indirectly via Josephson tunneling spectroscopy [52]. Our direct observation of spatially periodic gap variation in both single and bilayer cuprates exploring a large doping range adds strong support to the existence of pair density waves in them.

## Methods:

All spectra in the paper were obtained at 4K using standard lock-in techniques with $V_{bias}$=-100 mV, I = 100 pA and $V_{mod,rms}$ = 1.6 mV. The samples were cleaved in UHV environment at 77K and were quickly transferred to a custom built STM to ensure that the surface remains clean. The samples were studied using a tip cut from a Pt-Ir (80%-20%) wire.

## Acknowledgements:

The authors thank A. Tzalenchuk and S. Samaddar for helpful comments. The authors thank the Department of Physics of The Pennsylvania State University for generous support. In addition, R.B. also thanks the National Physical Laboratory, Teddington, UK for generous support.

# Supplementary Section

Section 1: Data on Bi-2212

In the main text, we presented data obtained on Bi-2201 showing the emergence of periodic gap variations extracted using our novel STUFF algorithm. In this supplementary section, we present results of running our analysis on a Bi-2212 sample. We see periodic gap variations, as shown in Fig. S1.

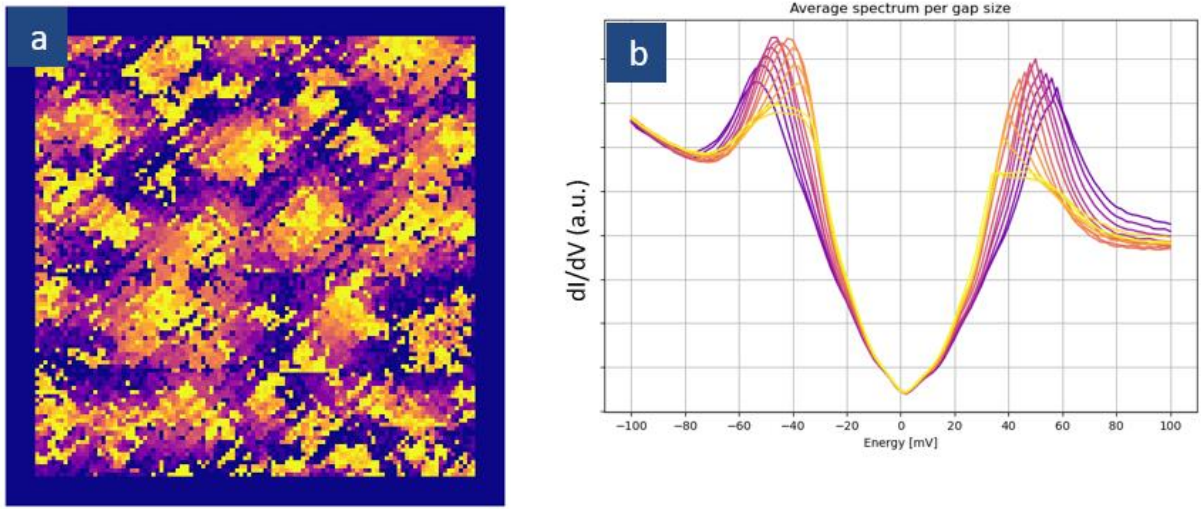

Figure S1: (a) Gap map obtained from STUFF algorithm for Bi-2212 ($T_c$=75K), with the spatially averaged spectra in the same color indexing scheme shown in (b). In (a), the field of view is 13.8 nm square split into 120 X 120 pixels. The color scale is the same as in Fig. 1b.

Section 2: Data on OPT35K and OD15K samples

Unlike the pronounced periodic gap variation discerned through STUFF on the underdoped Bi-2201 sample, elevating the doping level results in a diminished intensity of the pattern. In Figures S2 and S3 below, we present the result of our analysis on two more Bi-2201 samples: OPT35K and OD15K. As can be seen from the gap maps, the four-fold symmetry is gradually diminished as the sample doping is increased. The pattern weakens for the optimally doped sample and is barely perceptible on the OD15K sample.

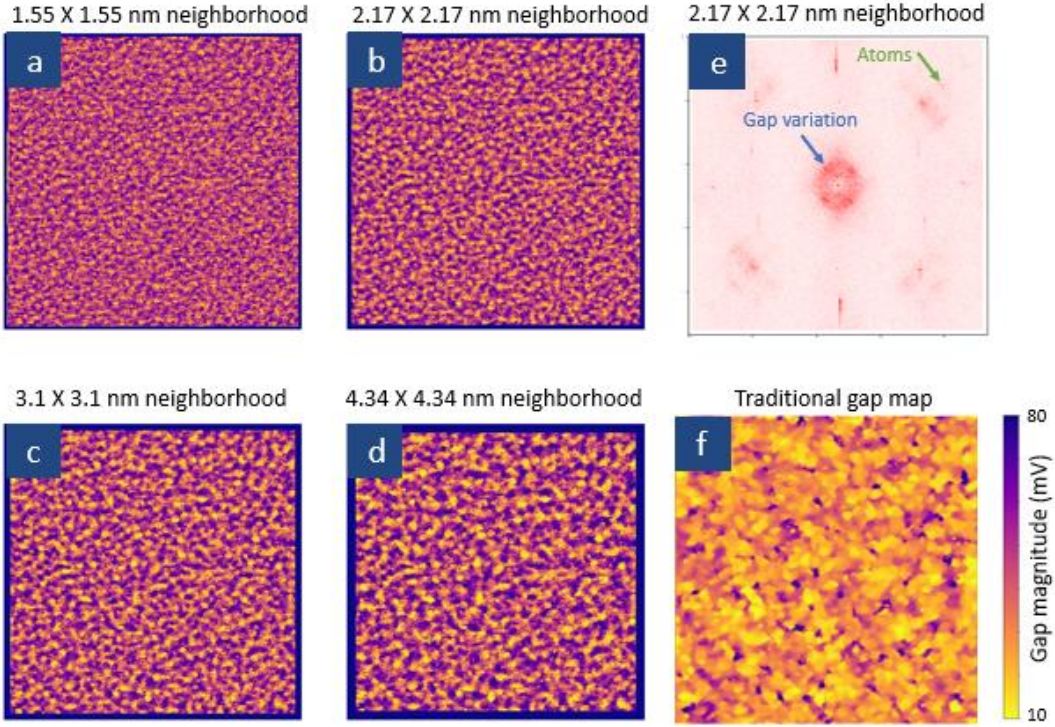

Figure S2: (a-d) Gap maps from using the STUFF algorithm on optimally doped (T$_c$=35K) Bi-2201 sample. The field of view is 72.5 nm square split into 468 X 468 pixels. Varying the neighborhood size, we see that the periodic gap variation pattern is significantly weaker than in the underdoped sample (T$_c$=32K), presented in the main text. (e) The weak periodicity can still be discerned in FFTs. (f) The traditional gap map, as usual, shows no clear pattern.

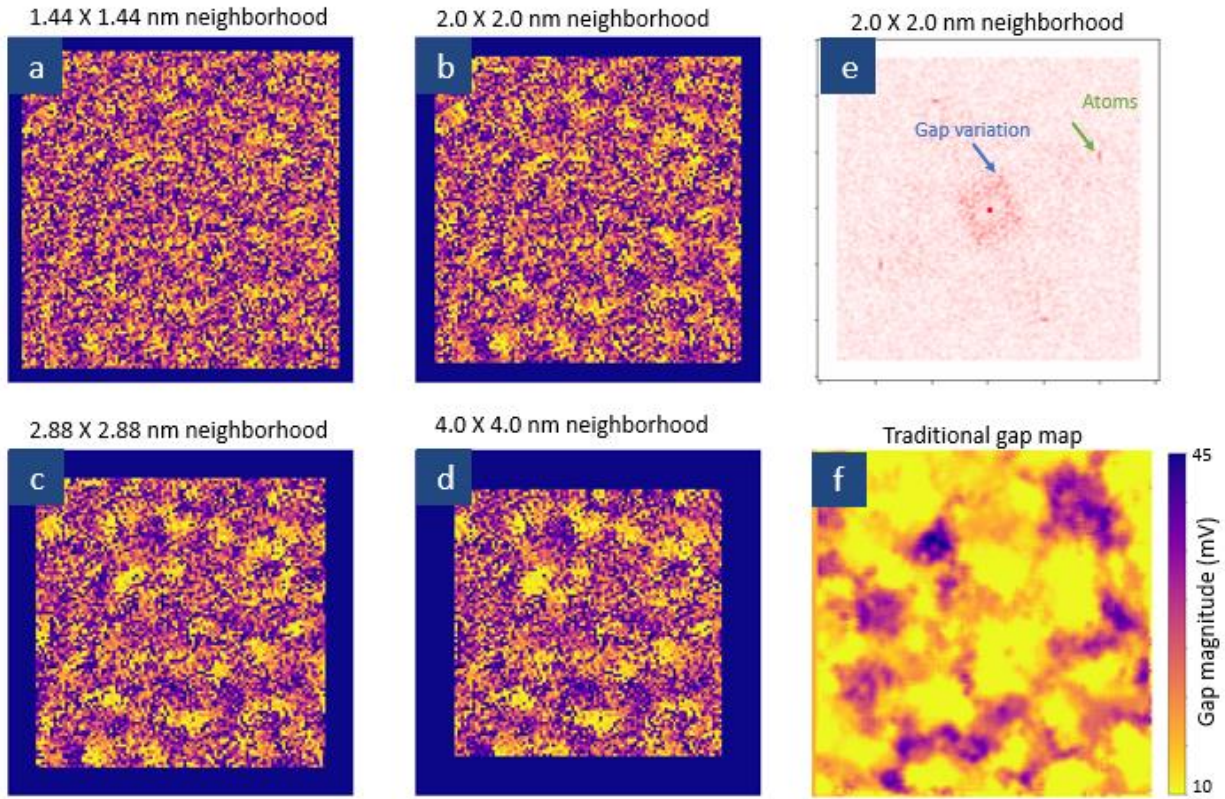

Figure S3: (a-d) Gap maps from using the STUFF algorithm on overdoped (T$_c$=15K) Bi-2201 sample. The field of view is 18.4 nm square split into 128 X 128 pixels. The periodic gap variation pattern is barely perceptible in this sample. (e) The weak periodicity discerned in FFTs. (f) The traditional gap map, as usual, shows no clear pattern.